\documentclass{appolb}
\usepackage{graphicx}

\begin{document}
\title{Measurement of the strong interaction induced shift and width of the 1s state of kaonic deuterium
at J-PARC 
\thanks{Acta Physica Polonica B}
}
\author{~
\address{~}
\\
{J. Zmeskal$^{1}$,
}
M. Sato$^{2}$, S. Ajimura$^{3}$, M. Bazzi$^{4}$, G. Beer$^{5}$, C. Berucci$^{1}$, H. Bhang$^{6}$, D. Bosnar$^{7}$, M. Bragadireanu$^{8}$, P. Buehler$^{1}$, L. Busso$^{9,10}$, M. Cargnelli$^{1}$, S. Choi$^{6}$, A. Clozza$^{4}$, C. Curceanu$^{4}$, A. D'uffizi$^{4}$, S. Enomoto$^{11}$, L. Fabbietti$^{12}$, D. Faso$^{9,10}$, C. Fiorini$^{13,14}$, H. Fujioka$^{15}$, F. Ghio$^{16}$, R. Golser$^{17}$, C. Guaraldo$^{4}$, T. Hashimoto$^{2}$, R.S. Hayano$^{18}$, T. Hiraiwa$^{3}$, M. Iio$^{11}$, M. Iliescu$^{4}$, K. Inoue$^{19}$, S. Ishimoto$^{11}$, T. Ishiwatari$^{20}$, K. Itahashi$^{2}$, M. Iwai$^{11}$, M. Iwasaki$^{2,21}$, S. Kawasaki$^{19}$, J. Lachner$^{17}$, P. Levi Sandri$^{4}$, Y. Ma$^{2}$, J. Marton$^{1}$, Y. Matsuda$^{22}$, Y. Mizoi$^{23}$, O. Morra$^{9}$, P. Moskal$^{24}$, T. Nagae$^{15}$, H. Noumi$^{3}$, H. Ohnishi$^{2}$, S. Okada$^{2}$, H. Outa$^{2}$, D. Pietreanu$^{8}$, K. Piscicchia$^{4,25}$, M. Poli Lener$^{4}$, A. Romero Vidal$^{26}$, Y. Sada$^{3}$, A. Sakaguchi$^{19}$, F. Sakuma$^{2}$, E. Sbardella$^{4}$, A. Scordo$^{4}$, M. Sekimoto$^{11}$, H. Shi$^{4}$, M. Silarski$^{4,24}$, D. Sirghi$^{4,8}$, F. Sirghi$^{4,8}$, K. Suzuki$^{1}$, S. Suzuki$^{11}$, T. Suzuki$^{18}$, K. Tanida$^{6}$, H. Tatsuno$^{11}$, M. Tokuda$^{21}$, A. Toyoda$^{11}$, I. Tucakovic$^{4}$, K. Tsukada$^{27}$, O. Vazquez Doce$^{12}$, E. Widmann$^{1}$, T. Yamaga$^{19}$, T. Yamazaki$^{2,18}$, Q. Zhang$^{2}$ 
\address{$^{1}$Stefan Meyer Institute for Subatomic Physics, Vienna, Austria\\
$^{2}$RIKEN Nishina Center, RIKEN, Wako, Japan\\
$^{3}$Research Center for Nuclear Physics (RCNP), Osaka University, Osaka, Japan\\
$^{4}$Laboratori Nazionali di Frascati dell’ INFN, Frascati, Italy\\
$^{5}$Department of Physics and Astronomy, University of Victoria, Victoria, Canada\\
$^{6}$Department of Physics, Seoul National University, Seoul, South Korea\\
$^{7}$Physics Department, University of Zagreb, Zagreb, Croatia\\
$^{8}$National Institute of Physics and Nuclear Engineering - IFIN HH, Romania\\
$^{9}$INFN Sezione di Torino, Torino, Italy\\
$^{10}$Dipartimento di Fisica Generale, Universita’ di Torino, Torino, Italy\\
$^{11}$High Energy Accelerator Research Organization (KEK), Tsukuba, Japan\\
$^{12}$Excellence Cluster Universe (TUM), Garching, Germany\\
$^{13}$Politecnico di Milano, Dipartimento di Elettronica, Milano, Italy\\
$^{14}$INFN Sezione di Milano, Milano, Italy \\
$^{15}$Department of Physics, Kyoto University, Kyoto, Japan\\
$^{16}$INFN Sezione di Roma I, Istituto Superiore di Sanit\'a, Roma, Italy\\
$^{17}$Isotope Research and Nuclear Physics, University of Vienna, Vienna, Austria\\
$^{18}$Department of Physics, University of Tokyo, Tokyo, Japan\\
$^{19}$Department of Physics, Osaka University, Osaka, Japan\\
$^{20}$Department of Physics, University of Vienna, Vienna, Austria\\
$^{21}$Department of Physics, Tokyo Institute of Technology, Tokyo, Japan\\
$^{22}$Graduate School of Arts and Sciences, University of Tokyo, Tokyo, Japan\\
$^{23}$Laboratory of Physics, Osaka Electro-Communication University, Osaka, Japan\\
$^{24}$Institute of Physics, Jagiellonian University, Cracow, Poland\\
$^{25}$Museo Storico della Fisica e Centro Studi e Ricerche Enrico Fermi, Roma, Italy\\
$^{26}$Universidade de Santiago de Compostela, Santiago de Compostela, Spain\\
$^{27}$Department of Physics, Tohoku University, Sendai, Japan\\}
}
\maketitle
\begin{abstract}
The antikaon-nucleon ($\overline{\hbox{K}}$N) interaction close to threshold provides crucial information on the interplay between spontaneous and explicit chiral symmetry breaking in low-energy QCD. In this context the importance of kaonic deuterium X-ray spectroscopy has been well recognized, but no experimental results have yet been obtained due to the difficulty of the measurement. 
We propose to measure the shift and width of the kaonic deuterium 1s state with an accuracy of 60 eV and 140 eV respectively at J-PARC. These results together with the kaonic hydrogen data (KpX at KEK, DEAR and SIDDHARTA at DA$\Phi$NE) will then permit the determination of values of both the isospin I=0 and I=1 antikaon-nucleon scattering lengths and will provide the most stringent constraints on the antikaon-nucleon interaction, promising a breakthrough.
Refined Monte Carlo studies were performed, including the investigation of  background suppression factors for the described setup. These studies have demonstrated the feasibility of determining the shift and width of the kaonic deuterium atom 1s state with the desired accuracy of 60 eV and 140 eV.
\end{abstract}

  
\section{Introduction}
\subsection{Physics motivation}
The antikaon-nucleon ($\overline{\hbox{K}}$N) interaction close to threshold provides crucial information on the interplay between spontaneous and explicit chiral symmetry breaking in low-energy QCD. Techniques such as scattering experiments and x-ray spectroscopy of kaonic atoms \cite{Zmeskal08} as well as production experiments close to threshold [2--7] have been used over many years. Theoretical investigations based on the experimental results have also been performed [8--19]. These approaches are complicated due to the presence of the $\Lambda$(1405) resonance located just below the K$^{-}$p threshold. At present, there are no lattice QCD calculations of antikaon-nucleon scattering lengths, although a theoretical framework has recently been proposed \cite{M. Lage}. In particular, the kaonic atom X-ray data provide the most precise values of the antikaon-nucleon scattering lengths at threshold.
The K$^{-}$p interaction is now well understood from the recent results of kaonic hydrogen obtained from KpX \cite{M. Iwasaki} at KEK, DEAR \cite{G. Beer1} and finally from SIDDHARTA at DA$\Phi$NE \cite{M. Bazzi1} along with theoretical calculations based on these results \cite{Y. Ikeda, M. Döring}. A first milestone was the KpX experiment which used fiducial volume cuts to suppress background, thus solving the long standing kaonic hydrogen puzzle. 
Although the importance of kaonic deuterium X-ray spectroscopy has been well recognized for more than 30 years (Dalitz et al., \cite{R.H. Dalitz}), no experimental results have yet been obtained due to the difficulty of the X-ray measurement. \textit{``The necessity to perform measurements of the kaonic deuterium ground state observables is justified by the fact that, unlike the case of pionic atoms, the measurement of only the kaonic hydrogen spectrum does not allow -- even in principle -- to extract independently both s-wave $\overline{\hbox{K}}$-nucleon scattering lengths $a_{0}$ and $a_{1}$''}, quoting from U.-G. Mei{\ss}ner \cite{U.‐G. Meißner2}. The kaonic deuterium X-ray measurement represents the most important experimental information missing in the field of the low-energy antikaon-nucleon interactions today. 
The kaonic hydrogen and deuterium data will also be indispensable when applying unitarized chiral perturbation theory to account for the $\Lambda$(1405) resonance in kaon-hadronic systems in nuclear matter. There are several hints of K$^{-}$ quasi-bound states in few-body nuclear systems by FINUDA \cite{Agnello} at DA$\Phi$NE and DISTO \cite{T. Yamazaki} at SATURNE and at J-PARC with the E15 experiment \cite{M. Iio} a search for a K$^{-}$pp bound system is ongoing, but the situation is not at all clear. 

\section{Present state of the art}
Recently, the most precise values of the strong interaction shift and width of the kaonic hydrogen 1s state were obtained by the SIDDHARTA experiment  \cite{M. Bazzi1}. The result of the pioneering KpX experiment (KEK-PS E228) together with those from DEAR and SIDDHARTA are shown in Fig. 1. In the latest SIDDHARTA kaonic hydrogen measurement at DA$\Phi$NE, an exploratory kaonic deuterium measurement was also made, in part to clarify the hydrogen background as well as to set a limit on the strength of the deuterium signal \cite{M. Bazzi2}.

\begin{figure}[htb]
\centerline{
\includegraphics[width=6.8cm]{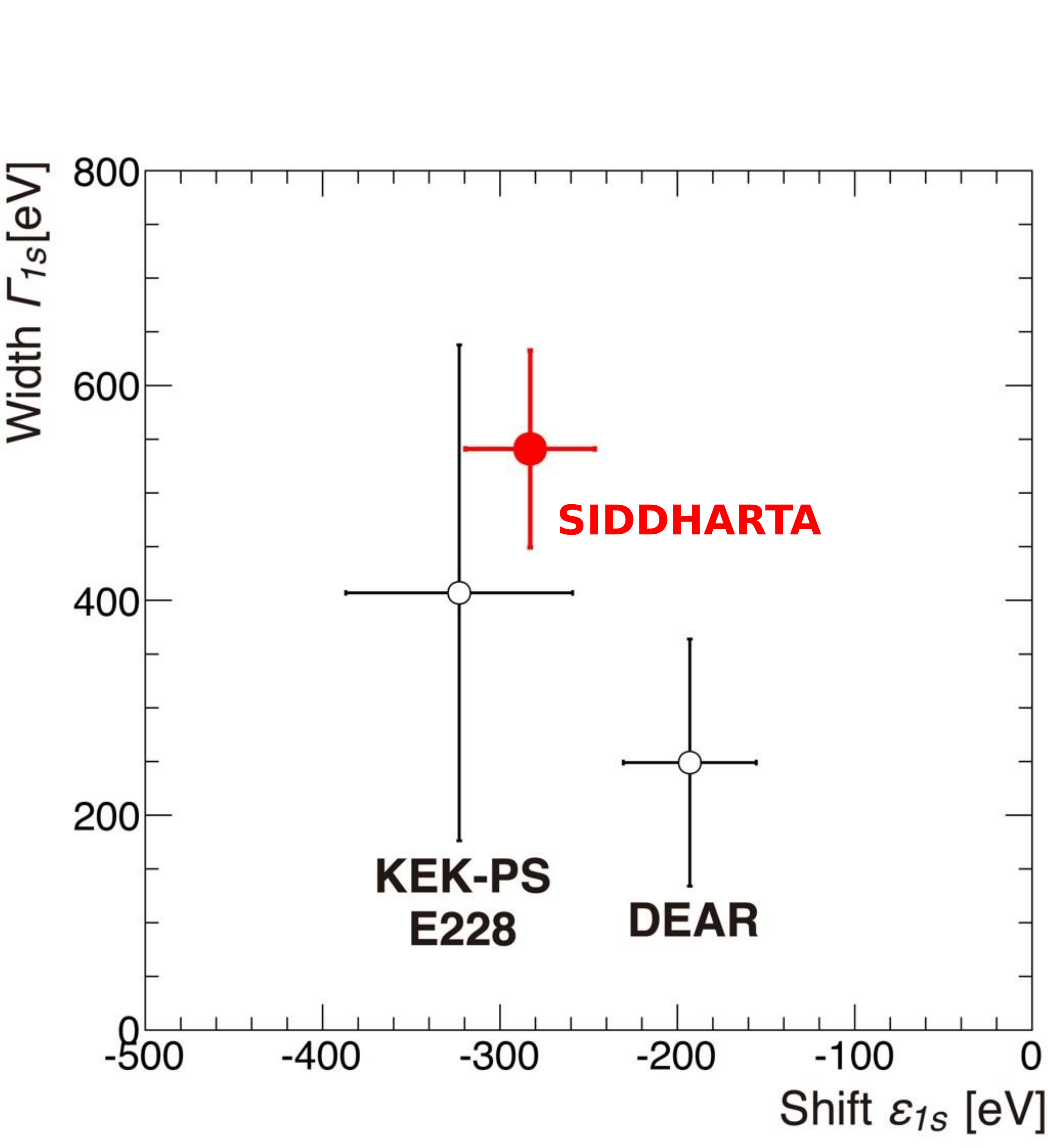}}
\caption{Comparison of experimental results for the strong interaction shift and width of the kaonic hydrogen 1s state.}
\label{Fig1.jpg}
\end{figure}
The experimentally determined shift and width are related to the s-wave scattering lengths at threshold \cite{Zmeskal08}. Because of isospin conservation, only the average value of the isospin I=0 and I=1 scattering lengths ($a_{0}$ and $a_{1}$) was obtained from the kaonic hydrogen measurement. In order to determine the isospin dependent scattering lengths, a measurement of the shift and width of the kaonic deuterium 1s state is definitely needed [24--27, 32--34].
On the theoretical side, there are many recent publications predicting quite consistent values of the shift and width for the kaonic deuterium 1s state (see Table 1).

\begin{center}
\begin{table}[htb]
\begin{center}
\begin{tabular}{cccc}
\hline
\hline
$a_{k-d}$ [fm]& ${\epsilon_{1s}}$ [eV]& ${\Gamma_{1s}}$ [eV]& ref.\\
\hline
$-$1.58+ i 1.37 & $-$887 & 757  & Mizutani 2013 \cite{T. Mizutani}\\
$-$1.48+ i 1.22 & $-$787 & 1011 & Schevchenko 2012 \cite{N.V. Shevchenko}\\
$-$1.46+ i 1.08 & $-$779 & 650  & Mei{\ss}ner 2011 \cite{M. Döring}\\
$-$1.42+ i 1.09 & $-$769 & 674  & Gal 2007 \cite{A. Gal}\\
$-$1.66+ i 1.28 & $-$884 & 665  & Mei{\ss}ner 2006 \cite{U.‐G. Meißner2}\\
\hline
\end{tabular}
\caption{Compilation of predicted K$^{-}$d scattering lengths $a_{k-d}$ and corresponding experimental quantities ${\epsilon_{1s}}$ and ${\Gamma_{1s}}$ (taken from \cite{M. Bazzi2}, Table1).}
\label{Table1}
\end{center}
\end{table}
\end{center}

\section{Proposed experimental method}

The proposed experiment will measure the transition X-ray energies to the ground state of kaonic deuterium atoms using recently developed Silicon Drift Detectors (SDDs) and using the experience gained with SDDs at KEK with the E570 \cite{S. Okada} experiment. In addition, charged particle tracking for background reduction will be used, as previously developed for the KpX experiment E228 \cite{M. Iwasaki} at KEK, as well as the background studies performed for the E17 experiment \cite{G. Beer2} at J-PARC to measure K$^{-}$$^{3}$He. Finally the results obtained from an exploratory measurement of kaonic deuterium with SIDDHARTA \cite{M. Bazzi2} were used to confine the values for the X-ray yield, shift and width used in the Monte Carlo studies for the experiment proposed in this article. 
The experimental challenge is the very small kaonic deuterium X-ray yield and the difficulty to perform X-ray spectroscopy in the high radiation environment of an extracted beam. It is therefore crucial to control and improve the signal-to-background ratio for a successful observation of the kaonic deuterium X-rays.

\section{The proposed experimental setup}

The proposed kaonic deuterium experiment will use the excellent features of the K1.8BR kaon beam line together with the K1.8BR spectrometer. We plan to upgrade the spectrometer with a cryogenic deuterium target surrounded by recently developed SDDs with a total area of 246 cm$^{2}$ (380 single elements) for X-ray detection.

\subsection{The setup at the K1.8BR spectrometer}

\begin{figure}[htb]
\centerline{
\includegraphics[width=9.5cm]{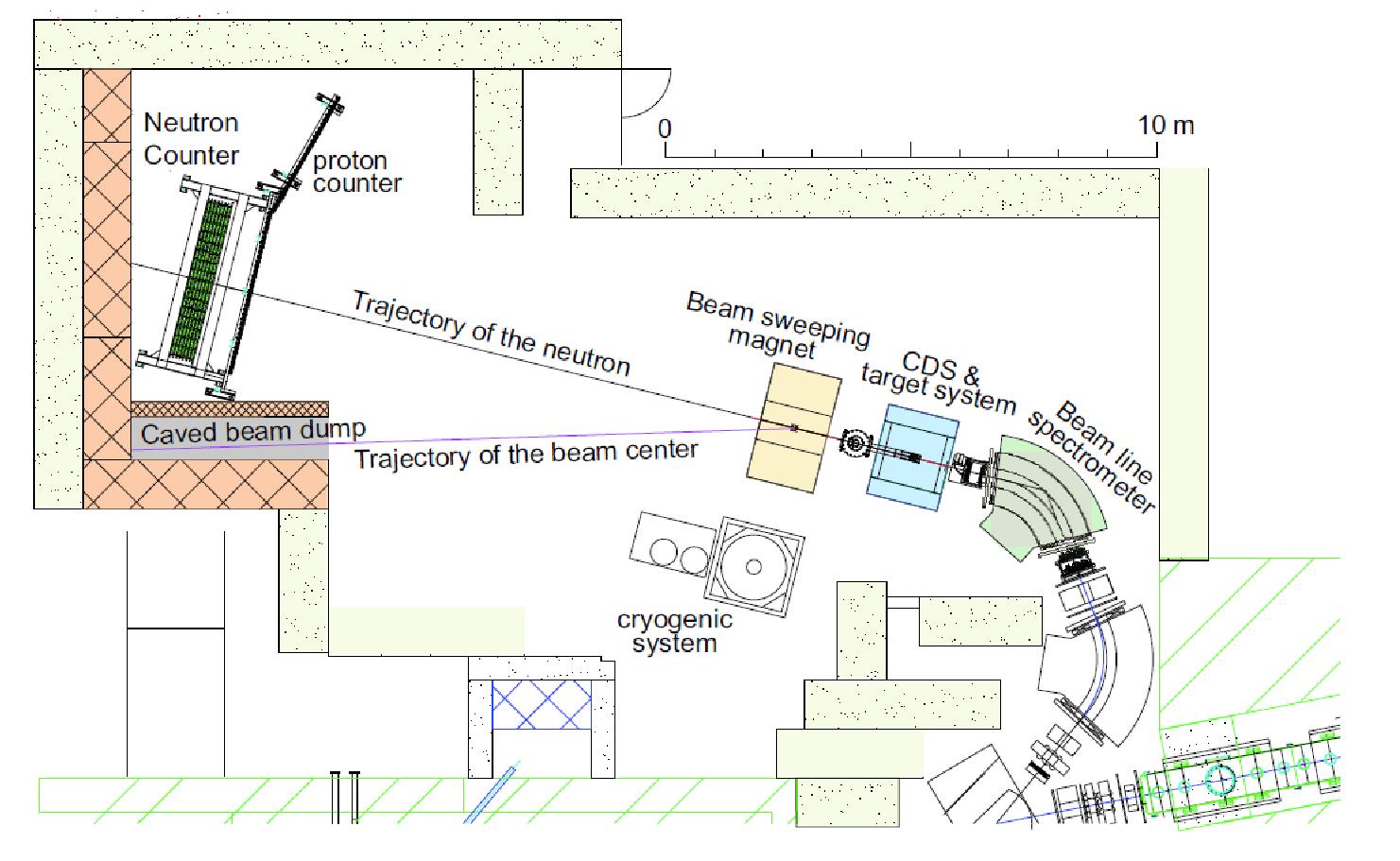}}
\caption{The J-PARC K1.8BR spectrometer. The setup consists of a beam line spectrometer, a cylindrical spectrometer system (CDS), a beam sweeping dipole magnet, a caved beam dump equipped with beam monitor hodoscopes, a neutron counter made of an array of plastic scintillation counters equipped with charged veto counters and a proton counter hodoscope.}
\label{Fig2.jpg}
\end{figure}

The K1.8BR multi-purpose spectrometer \cite{Agari} (see Fig. 2) has a quite unique features for our needs, namely a large acceptance cylindrical spectrometer system (CDS), consisting of a cylindrical drift chamber (CDC) for charged particle tracking, essential for efficient background reduction for the proposed study of kaonic deuterium X-rays. The cylindrical drift chamber, with a diameter of 1060 mm and length of 950 mm, is surrounded by a cylindrical detector hodoscope (CDH) to trigger on decay particles. 
These parts of the experimental setup will be utilized for the proposed experiment. They will be upgraded with a cryogenic target system and a novel X-ray detector system with a total area of 246 cm$^{2}$, having excellent energy resolution (better than 150 eV at 6 keV) and improved timing ($\sim$100 ns).
The proposed setup (Fig. 3) will allow tracking of the incoming kaons with segmented plastic scintillators -- used also as start counters (T1, T0) -- and the beam line chamber (BLC).  The charged particles produced due to kaon absorption on the nucleus will be tracked by the large cylindrical drift chamber (CDC), while the cylindrical detector hodoscope (CDH) is used as the trigger in coincidence with the start counters (T1, T0).

\begin{figure}[htb]
\centerline{
\includegraphics[width=6.cm]{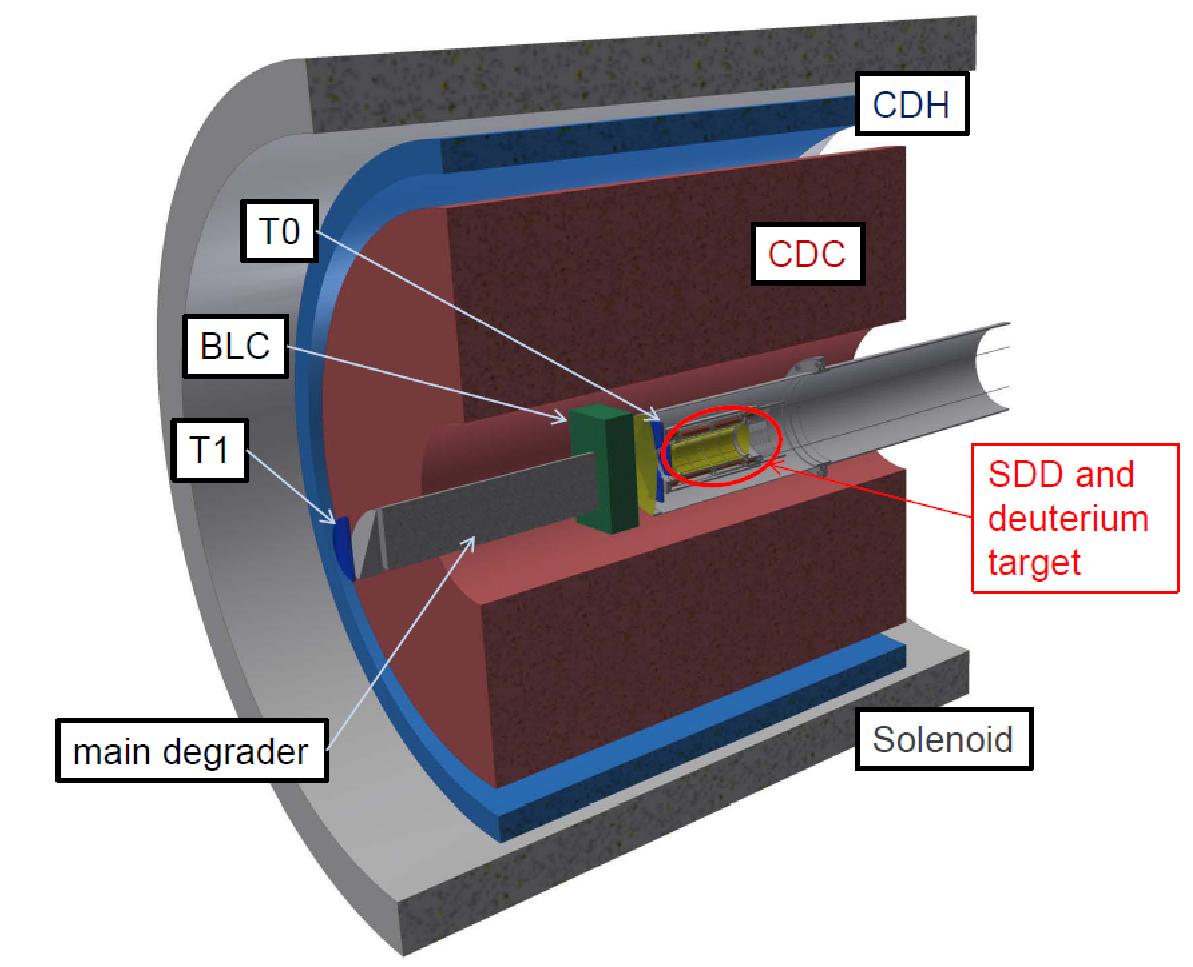}}
\caption{Sketch of the proposed setup for the kaonic deuterium measurement.}
\label{Fig3.jpg}
\end{figure}

\subsection{New Silicon Drift Detectors (SDDs)}

We plan to use new SDD chips, which were developed in collaboration with Fondazione Bruno Kessler FBK and Politecnico di Milano in Italy.

\begin{figure}[htb]
\centerline{
\includegraphics[width=8.cm]{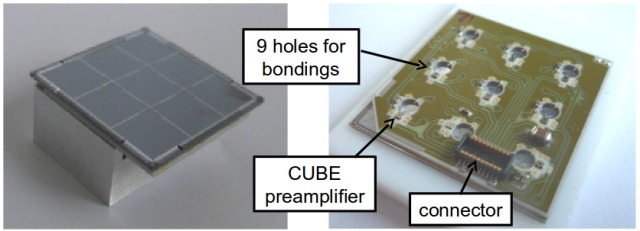}}
\caption{Monolithic array of  3x3 SDDs each 8x8 mm$^{2}$ (left); backside of the SDD array mounted on a printed circuit board with preamplifier chip (CUBE) for detector segment (right).}
\label{Fig4.jpg}
\end{figure}

Monolithic arrays with 9 detector segments with a total area of 5.76 cm$^{2}$ (see Fig. 4) are ideally suited for the proposed experiment. For these SDDs, special preamplifier chips (CUBE), were developed. 
First test measurements were performed successfully. For example, a 72 hour stability test, performed at an SDD temperature of 120 K, achieved an energy resolution of 130 eV at 6 keV (see Fig. 5). The drift time at this working temperature was measured to be around 200 ns.

\begin{figure}[htb]
\centerline{
\includegraphics[width=6.cm]{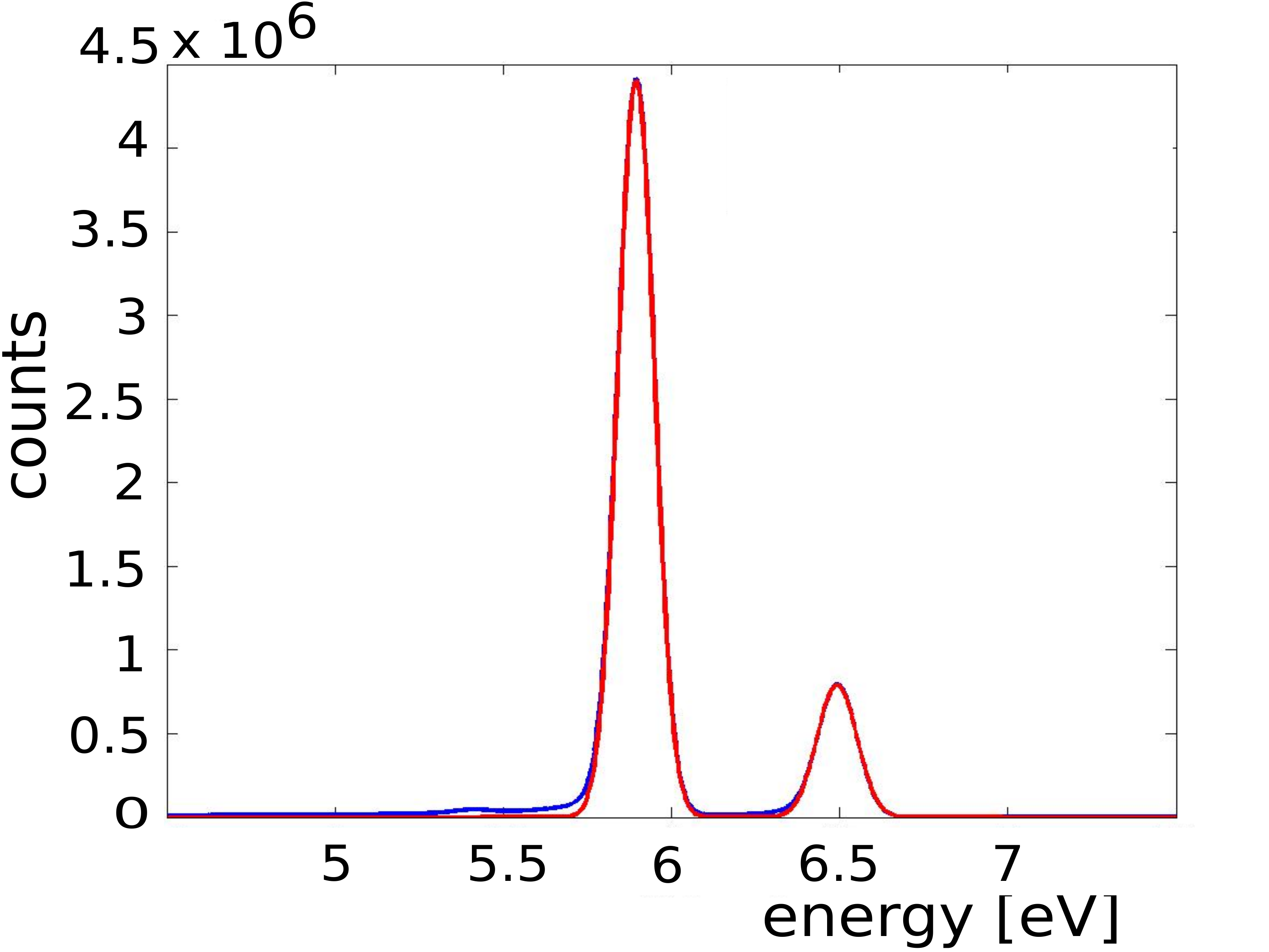}}
\caption{Test of a single cell at a temperature of 100 K, measured for 72 hours using a Fe-55 radioisotope, which decays via electron capture to Mn-55. The $MnK_{\alpha}$-line at 5.9 keV, emitted after electron capture, is measured with high resolution, leading to a FWHM of 127 eV.}
\label{Fig5.jpg}
\end{figure}

\subsection{Cryogenic deuterium target and SDD mounting structure}

The cryogenic target cell will be made of a 75µ${\mu}$m Kapton body with a diameter of 65 mm and a length of 160 mm, with reinforcement structure made out of aluminium. In addition, the cooling and mounting structure of the SDDs is used, in addition, to reinforce the target cell in the longitudinal direction. The working temperature of the target cell is around 25 K with a maximum pressure of 0.35 MPa. With these parameters, a gas density of 5\% liquid deuterium density (LDD) will be achieved.
Finally, 48 monolithic SDD arrays will be placed close together around the target, with a total area of 246 cm$^{2}$ containing 380 readout channels (see Fig. 6).

\begin{figure}[htb]
\centerline{
\includegraphics[width=7.cm]{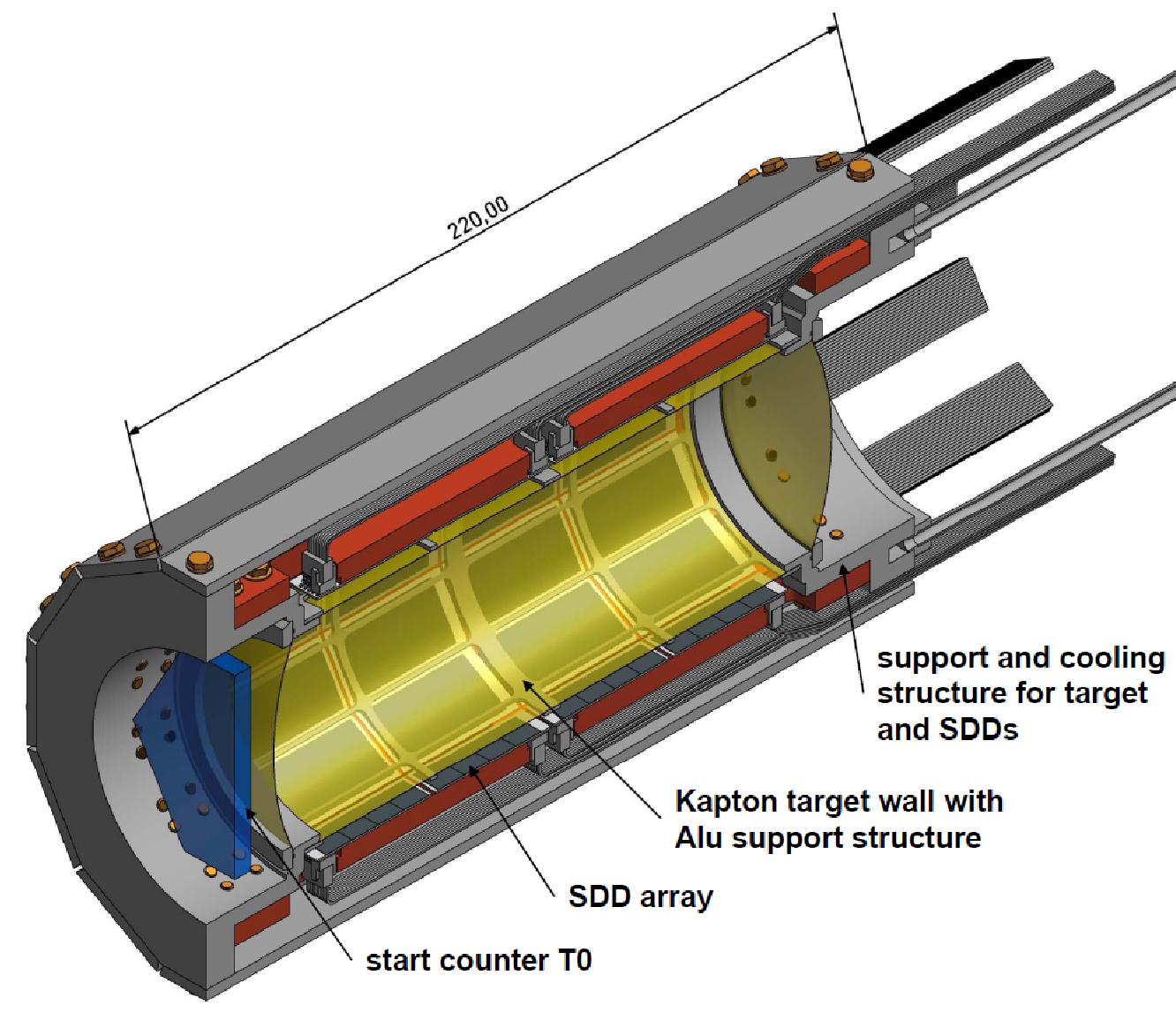}}
\caption{Design of the cryogenic target and X-ray detector system. The target cell, with a diameter of 65 mm and a length of 160 mm, is closely surrounded by SDDs, about 5 mm away from the target wall.}
\label{Fig6.jpg}
\end{figure}

\section{Monte Carlo studies}
\subsection{Degrader optimisation, kaon stopping distribution}

The kaon beam properties were taken from a measurement at the K1.8 BR in June 2012 with a kaon momentum of 1000 MeV/c.
For use as a kaon degrader, several materials (carbon, polyethylene and iron) were compared in a simulation for kaons with a central momentum of 700 MeV/c. The highest stopping rates were obtained with a carbon degrader of about 40 cm thickness.
In Table 2 the optimised results for the carbon degrader thicknesses are summarised for gas and liquid targets. An additional prism-shaped degrader was used to compensate for the position dependence of the momentum. The results of the kaon stopping optimisation in gaseous and liquid deuterium targets respectively is shown in Fig. 7. The simulation to optimise the kaon stops in deuterium started with $7\cdot10^{6}$ K$^{-}$ and a kaon momentum of 0.7 GeV/c.
All further calculations are performed only for a gaseous target with a density of 5\% LDD.
\begin{center}
\begin{table}[htb]
\begin{center}
\begin{tabular}{ccc}
\hline
\hline
Degrader & Kaon target stops  & Target density\\
thickness [cm]  & per beam kaon (10$^{-3}$)  &                  \\
\hline
40 & 0.31  & 0.03\\
40 & 0.50  & 0.05\\
39 & 9.5  & 1\\
\hline
\end{tabular}
\caption{Kaon stopping density in gaseous and liquid targets, optimised for carbon degrader. The target density is given relative to the liquid deuterium density (LDD).}
\label{Table2}
\end{center}
\end{table}
\end{center}

\begin{figure}[htb]
\centerline{
\includegraphics[width=10.5cm]{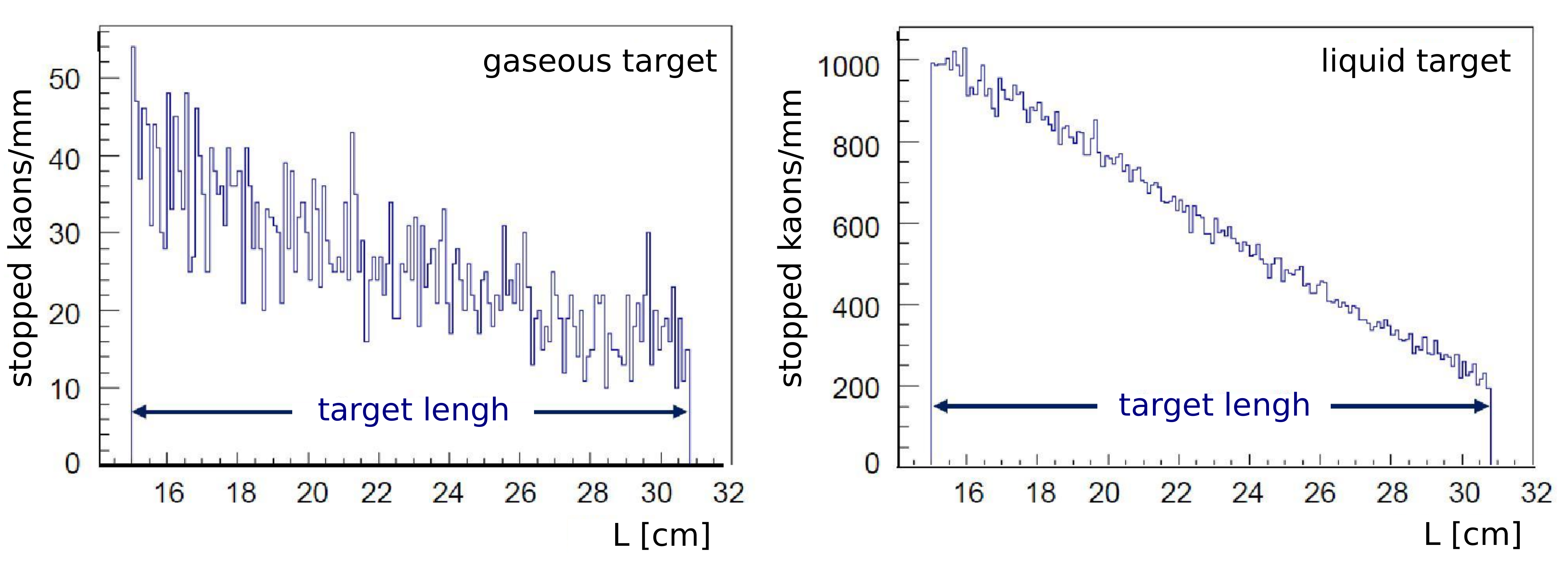}}
\caption{Kaon stopping distribution in the deuterium gas target with a density of 5\% LDD (left), and in the liquid deuterium target (right).}
\label{Fig7.jpg}
\end{figure}

\subsection{X-ray yields, cascade calculations}

The most recent cascade calculations for kaonic deuterium are shown in Fig. 8, where the X-ray yields per stopped kaon for $K_{\alpha}$, $K_{\beta}$ and $K_{\gamma}$ transitions over a wide density range were calculated.
As input in our Monte Carlo simulation we used a $K_{\alpha}$ yield of 0.1\% for the gas targets and 0.01\% for the liquid target, which are between the calculated values of Koike \cite{T.Koike} and Jensen \cite{T. S. Jensen}.

\begin{figure}[htb]
\centerline{
\includegraphics[width=10.5cm]{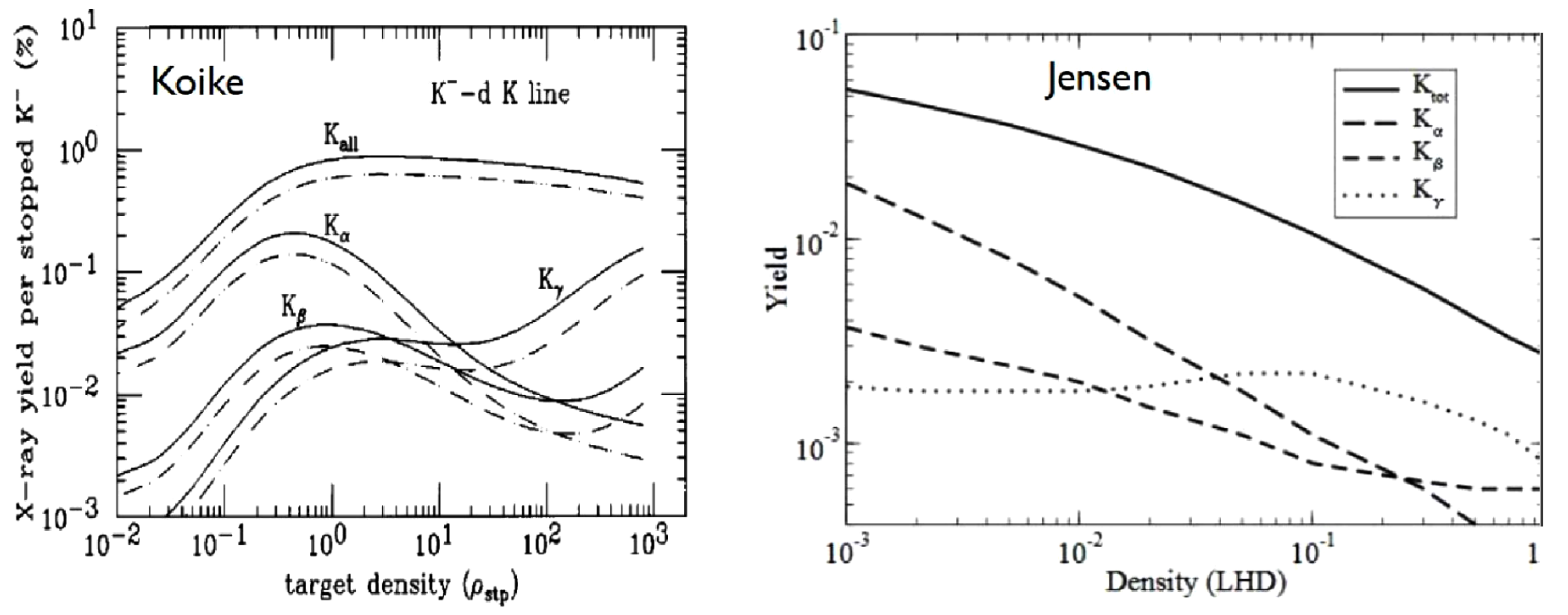}}
\caption{Kaonic deuterium cascade calculations, for the X-ray yield of $K_{\alpha}$, $K_{\beta}$, $K_{\gamma}$ and $K_{tot}$; figure from reference \cite{T.Koike} (left) and from \cite{T. S. Jensen} (right).}
\label{Fig8.jpg}
\end{figure}

\subsection{Feasibility study}

We have simulated the kaonic deuterium X-ray spectrum ($3\cdot10^{8}$ K$^{-}$ produced per day) using the new X-ray detector with an active area of 246 cm$^{2}$. The gaseous target density is 5\% of LDD. The results of this calculation are summarised in Table 3.

\begin{center}
\begin{table}[htb]
\begin{center}
\begin{tabular}{lc}
\hline
\hline
produced K$^{-}$ per day (80\% duty cycle) & $3.0\cdot10^{8}$\\
numbers of K$^{-}$ at beam counter (T1) & $17.4\cdot10^{6}$\\
trigger: T0$\otimes$(charged particle in CDH) & $11.4\cdot10^{6}$\\
number of gas stops & $1.5\cdot10^{5}$\\ 
total synchronous BG per keV at 7 keV & $3.6\cdot10^{4}$\\
synchronous BG per keV at 7 keV with & 90\\
fiducial cut and charged particle veto & \\
asynchronous BG per keV at 7 keV & 30\\
detected $K_{\alpha}$ events for a yield $Y(K\alpha)$=0.1\%, & 30\\
with fiducial cut and charged particle veto &  \\
\hline
\end{tabular}
\caption{Simulated values for synchronous background and the $K_{\alpha}$ transition X-ray yield. For the asynchronous background the adapted estimation of E17 is used. Background and signal are calculated for an active SDD area of 246 cm$^{2}$.}
\label{Table3}
\end{center}
\end{table}
\end{center}

The yield ratios of the $K_{\alpha}$:$K_{\beta}$:$K_{total}$ transitions were taken from the kaonic hydrogen data, but with an assumed $K_{\alpha}$ yield of 10$^{-3}$ for a gaseous deuterium target. For the strong interaction induced shift and width, theoretical predictions of shift = $-$800 eV and width of 750 eV were used (see Table 1).
The simulated spectrum for the transition energies of kaonic deuterium atoms is shown in Fig. 9, using the vertex cut method (successfully used with E570 at KEK and planned as well for E17 at J-PARC) with a defined volume inside the target (5 mm away from the walls), and a charged particle veto for tracks passing through or nearby to the SDDs.
\begin{figure}[htb]
\centerline{
\includegraphics[width=8.cm]{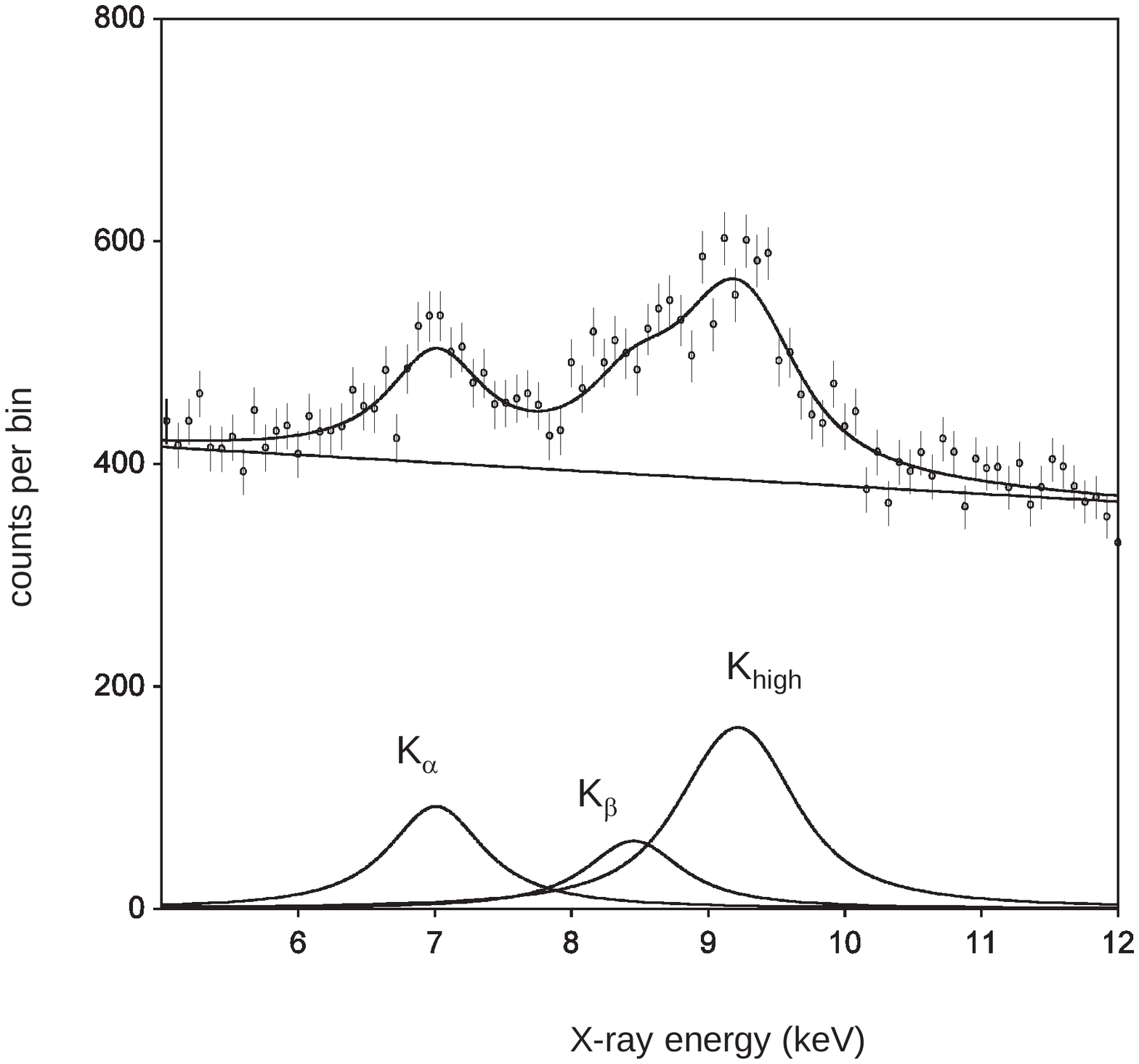}}
\caption{Simulated kaonic deuterium x-ray spectrum, assuming 10$^{10}$ produced K$^{-}$ and a detector area of 246 cm$^{2}$. In addition demanding a vertex cut (fiducial volume 5 mm off the walls) as well as the charged particle veto for tracks passing through the SDDs.}
\label{Fig9.jpg}
\end{figure}
For a gas density of 5\%, the estimated signal to background ratio (integral) is 1:4 (see Fig. 9). Fitting a set of simulated data, we extracted the shift and width with a precision of 60 eV and 140 eV, respectively, which is comparable with the precision of the kaonic hydrogen (K$^{-}$p) SIDDHARTA result of the shift and width measurement.

\section{Conclusion}

Assuming a kaon intensity of about $3\cdot10^{8}$ K$^{-}$ per day with a momentum of 700 MeV/c and rejecting the kaon correlated background using various cuts, it will be possible to collect 1000 $K_{\alpha}$ events with a signal to background ratio of 1:4 in 30 days. The asynchronous background will be efficiently reduced due to the unprecedented timing resolution of the new SDD ($\sim$100 ns). 

\section*{Acknowledgements}

This work is supported by RIKEN, KEK, RCNP, a Grant-in-Aid for Scientific Research on Priority Areas [No.17070005 and No.20840047], a Grant-in-Aid for Specially Promoted Research [No.20002003], a Grant-in-Aid for Young Scientists (Start-up) [No.20028011], a Grant-in-Aid for Scientific Research on Innovative Areas [No.21105003], and the Austrian Science Fund (FWF) [P24756-N20].



\begin{thebibliography}{999}
\vspace*{-0.25cm}
\bibitem{Zmeskal08} J. Zmeskal, Progr. Part. Nucl. Phys. 61 (2008) 512.
\bibitem{P.Winter} P.Winter, et al., Phys. Lett. B 635 (2006) 23-29.
\bibitem{Y. Maeda} Y. Maeda, et al., Phys. Rev. C 77 (2008) 015204.
\bibitem{M. Silarski} M. Silarski, et al., Phys. Rev. C 80 (2009) 045202.
\bibitem{Q. J. Ye} Q. J. Ye, et al., Phys. Rev. C 85 (2012) 035211.
\bibitem{P. Moskal} M. Silarski, P. Moskal, Phys. Rev. C 88 (2013) 025205.
\bibitem{G.Agakishiev} G.Agakishiev, et al., Phys. Rev. C 87 (2013) 025201.
\bibitem{U.‐G. Meißner1} U.-G. Mei{\ss}ner, U. Raha, A. Rusetsky, Eur. Phys. J. C 35 (2004) 349.
\bibitem{B. Borasoy1} B. Borasoy, R. Ni{\ss}ler, W. Weise,Phys. Rev. Lett. 94 (2005) 213401.
\bibitem{B. Borasoy2} B. Borasoy, R. Ni{\ss}ler, W. Weise, Eur. Phys. J. A 25 (2005) 79.
\bibitem{B. Borasoy3} B. Borasoy,  U.-G. Mei{\ss}ner, R. Ni{\ss}ler, Phys. Rev. C74 (2006) 055201.
\bibitem{J.A. Oller1} J.A. Oller, J. Prades, M. Verbeni, Phys. Rev. Lett. 95 (2005) 172502.
\bibitem{J.A. Oller2} J.A. Oller, Eur. Phys. J. A 28 (2006) 63.
\bibitem{B. Borasoy4} B. Borasoy, R. Ni{\ss}ler, W. Weise, Phys. Rev. Lett. 96 (2006) 199201.
\bibitem{J.A. Oller3} J.A. Oller, J. Prades, M. Verbeni, Phys. Rev. Lett. 96 (2006) 199202.
\bibitem{J. Révai} J. R{\'e}vai, N.V. Shevchenko, Phys. Rev. C 79, 035202 (2009).
\bibitem{A. Cieplý} A. Ciepl{\'y}, J. Smejkal, Eur. Phys. J. A34 (2007) 237.
\bibitem{E. Oset} E. Oset, A. Ramos, Nucl. Phys. A 635 (1998) 99.
\bibitem{W. Weise} W. Weise, Nucl. Phys. A. 835 (2010) 51.
\bibitem{M. Lage} M. Lage, U.-G. Mei{\ss}ner, A. Rusetsky, Phys. Lett. B 681 (2009) 439.
\bibitem{M. Iwasaki} M. Iwasaki, et al., Phys. Rev. Lett. 78 (1997) 3067.
\bibitem{G. Beer1} G. Beer, et al., Phys. Rev. Lett. 94 (2005) 212302.
\bibitem{M. Bazzi1} M. Bazzi, et al., Phys. Lett. B 704 (2011) 113.
\bibitem{Y. Ikeda} Y. Ikeda, T. Hyodo, W. Weise, Phys. Lett. B 706 (2011) 63.
\bibitem{M. Döring} M. D{\"o}ring, U.‐G. Mei{\ss}ner, Phys. Lett. B 704 (2011) 663.
\bibitem{R.H. Dalitz} R.H. Dalitz, J. McGinley, C. Belyea, S. Anthony, ed. by B. Povh (M.P.I., Heidelberg 1982), p. 201.
\bibitem{U.‐G. Meißner2} U.-G. Mei{\ss}ner, U. Raha, A. Rusetsky, Eur. Phys. J. C 47 (2006) 473.
\bibitem{Agnello} Agnello, et al., Phys. Rev. Lett. 94 (2005) 212303.
\bibitem{T. Yamazaki} T. Yamazaki, et al., Phys. Rev. Lett. 104 (2010) 132502.
\bibitem{M. Iio} M. Iio , et al., Proposals for Nuclear and Particle Physics Experiments at J-PARC, the 1st PAC meeting, Fri 30 June - Sun 02 July, 2006.
\bibitem{M. Bazzi2} M. Bazzi, et al., Nucl. Phys. A 907 (2013) 69.
\bibitem{S. Bianco} S. Bianco, et al., La Rivista del Nuovo Cimento vol.22 no. 11 (1999) 1.
\bibitem{A. Gal} A. Gal, Int. J. Mod. Phys. A 22 (2007) 226.
\bibitem{V. Baru} V. Baru, E. Epelbaum, A. Rusetsky, Eur. Phys. J. A. 42 (2009) 111.
\bibitem{T. Mizutani} T. Mizutani, C. Fayard, B. Saghai, K. Tsushima, arXiv:1211.5824[hep-ph] (2013).
\bibitem{N.V. Shevchenko} N.V. Shevchenko, Nucl. Phys. A 890-891 (2012) 50-61.
\bibitem{S. Okada} S. Okada, et al., Phys. Lett. B 653 (2007) 387.
\bibitem{G. Beer2} G. Beer, et al., Proposals for Nuclear and Particle Physics Experiments at J-PARC, the 1st PAC meeting, Fri 30 June - Sun 02 July, 2006.
\bibitem{Agari} K.Agari, et al., Progress of Theoretical and Experimental Physics (PTEP) 2012, 02B011.
\bibitem{T.Koike} T.Koike, T. Harada, Y. Akaishi, Phys. Rev. C 53 (1996) 79.
\bibitem{T. S. Jensen} T. S. Jensen, Proceedings of DAFNE 2004: Physics at meson factories, June 2004.
\end{thebibliography}
\end{document}